# How the IMF $B_y$ Induces a Local $B_y$ Component During Northward IMF $B_z$ and Characteristic Timescales

P. Tenfjord[1], N. Østgaard[1], S. Haaland[1,2], K. Snekvik[1], K. M. Laundal[1], J. P. Reistad[1], R. Strangeway[3], S. E. Milan[1,4], M. Hesse[1], and A. Ohma[1]

[1]Birkeland Centre for Space Science, Department of Physics and Technology, University of Bergen, Bergen, Norway, [2]Max Planck Institute for Solar System Research, Göttingen, Germany, [3]Institute of Geophysics and Planetary Physics, University of California, Los Angeles, CA, USA, [4]Department of Physics and Astronomy, University of Leicester, Leicester, UK

**Abstract** We use the Lyon-Fedder-Mobarry global magnetohydrodynamics model to study the effects of the interplanetary magnetic field (IMF) $B_y$ component on the coupling between the solar wind and magnetosphere-ionosphere system when IMF $B_z > 0$. We describe the evolution of how a magnetospheric $B_y$ component is induced on closed field lines during these conditions. Starting from dayside lobe reconnection, the magnetic tension on newly reconnected field lines redistribute the open flux asymmetrically between the two hemispheres. This results in asymmetric magnetic energy density in the lobes. Shear flows are induced to restore equilibrium, and these flows are what effectively induces a local $B_y$ component. We show the radial dependence of the induced $B_y$ and compare the results to the induced $B_y$ during southward IMF conditions. We also show the response and reconfiguration time of the inner magnetosphere to IMF $B_y$ reversals during northward IMF $B_z$. A superposed epoch analysis of magnetic field measurements from seven Geostationary Operational Environmental Satellite spacecraft at different local times both for negative-to-positive and positive-to-negative IMF $B_y$ reversals is presented. We find that the induced $B_y$ responds within 16 min of the arrival of IMF $B_y$ at the bow shock, and it completely reconfigures within 47 min.

## 1. Introduction

The large-scale magnetospheric dynamics is primarily driven by dayside reconnection between the terrestrial magnetic field and the interplanetary magnetic field (IMF). When the IMF is oriented southward, it reconnects with closed magnetospheric field lines at low latitudes, and subsequently opened field lines are dragged across the polar caps into the magnetosphere by the solar wind (Dungey, 1961). Over long timescales, dayside reconnection is balanced by magnetotail reconnection.

When the IMF is directed northward, dayside reconnection primarily takes place between the IMF and lobe field lines at or behind the cusps (Russell, 1972; Watanabe et al., 2005). This is sometimes called high-latitude reconnection due to the fact that the magnetospheric field lines map to high latitudes in the ionosphere. It is also termed lobe reconnection, since the newly reconnected field lines were previously part of the magnetospheric lobe field (Crooker, 1992).

In the event in which northward IMF $B_z$ dominates, the same interplanetary field line might reconnect in both hemispheres. This is known as dual-lobe reconnection. The process describes a situation where open lobe field lines are replaced by closed field lines formed at the dayside. In theory, it is possible that dual-lobe reconnection, if proceeding over a significant period of time, leads to a situation where the entire magnetosphere consists of closed field lines. The incoming IMF field line does not necessarily reconnect at the two cusps simultaneously but may reconnect first in one hemisphere due to either IMF $B_x$ or dipole tilt (Cowley, 1981; Imber et al., 2006; Song & Russell, 1992).

Single-lobe reconnection refers to the situation when the IMF reconnects only in one hemisphere, such that the newly reconnected flux tube has only one foot point on Earth and the other out in the solar wind. Single-lobe reconnection is often associated with an IMF $B_y$ component (Fuselier et al., 2014; Zhong et al., 2013). Single-lobe reconnection can take place in both hemispheres simultaneously, but the evolution and dynamics are controlled by the forces acting on that field line and thus independent from the opposite hemisphere (Østgaard et al., 2005).

The ionospheric convection associated with purely northward IMF is characterized by two lobe convection cells (e.g., Crooker, 1979; Förster et al., 2008; Potemra et al., 1984; Wilder et al., 2009). Here the reconnection foot point starts at high latitudes. The initial sunward convection occurs as the new closed field line "sinks" into the magnetoshere, due to the sunward tension on the newly reconnected field line (Song & Russell, 1992). The foot point convection then deflects toward the dusk or dawn, depending on which side of the noon-midnight meridian it is initially located on. This is due to magnetospheric azimuthal motion resulting from either the interchange instability dispersing the flux tube along the magnetopause to form a boundary layer (Song & Russell, 1992), accumulation of magnetic pressure forcing azimuthal circulation, or magnetic tension related to the IMF $B_y$ component. The last portion of the ionospheric convection completing the "circle" represents transport of fresh flux from the lobe field out to the region where reconnection occurs, due to a rarefaction wave from the reconnection region.

In the presence of an IMF $B_y$ component, it is likely that both single- and dual-lobe reconnection occurs. A smaller clock angle favors dual-lobe reconnection. The reconnection region, determined by antiparallel geometry (or more advanced reconnection models, e.g., Fuselier et al., 2014; Trattner et al., 2004), is displaced toward the dawnside or duskside, depending on the orientation of the IMF $B_y$. The incoming field lines reconnecting close to the noon-midnight meridian are more likely to undergo dual-lobe reconnection, compared to the field lines reconnecting on the dawnward or duskward side of this region.

The effects of IMF $B_y$ on the ionospheric flow are characterized by a departure from the symmetric two-cell configuration to the growth of one of the lobe cells, while the other will contract (Cumnock et al., 1992; Huang et al., 2000; Reiff & Burch, 1985). Statistical maps of the ionospheric convection can be found in Förster et al. (2008).

During periods of prolonged northward directed IMF, a bright auroral spot can be observed at the foot point of merging field lines. This cusp spot is located poleward of the main dayside auroral oval and is observed during both single- and dual-lobe reconnection (Imber et al., 2006; Milan et al., 2000; Østgaard et al., 2005; Phan et al., 2003). Under the same condition, that is, sustained northward IMF, theta aurora or transpolar arcs are commonly observed. Both the location and motion of these transpolar arcs are observed to depend on the IMF $B_y$ component (Fear & Milan, 2012; Fear et al., 2015; Milan et al., 2005; Tanaka et al., 2004). Østgaard et al. (2003) showed two cases where the theta aurora was only seen in one hemisphere, which was also attributed to IMF $B_y$. As suggested by Østgaard et al. (2003), it also indicates that single-lobe reconnection can occur in one hemisphere only due to a strong IMF $B_x$ component.

Understanding the effect of IMF $B_y$ during northward directed IMF, and how this induces a local $B_y$ component in the magnetosphere, is therefore paramount to fully understand the coupled solar wind-magnetosphere-ionosphere system.

Empirical studies have shown that an induced $B_y$ component inside the magnetopause is present during northward IMF with a significant $B_y$ component (Cao et al., 2014; Cowley & Hughes, 1983; Kaymaz et al., 1994; Rong et al., 2015; Wing et al., 1995). Observations also suggest that the induced $B_y$ is larger during northward IMF $B_z$ compared to southward (Cowley & Hughes, 1983).

How a $B_y$ component is established in the closed magnetosphere has been a controversy. It has often been referred to as simple penetration of IMF $B_y$ (Kozlovsky, 2003; Rong et al., 2015), while others have suggested that $B_y$ is transported into the closed magnetosphere through tail reconnection (Fear & Milan, 2012; Grocott et al., 2007; Hau & Erickson, 1995; Østgaard et al., 2004; Stenbaek-Nielsen & Otto, 1997). However, as shown in Tenfjord et al. (2015), the IMF-induced observed $B_y$-related asymmetries in the magnetosphere may be explained by asymmetric addition of magnetic flux to the tail lobes about the noon-midnight meridian (Cowley, 1981; Khurana et al., 1996). Different mechanisms can generate the induced $B_y$ component (see Petrukovich, 2011; Tenfjord et al., 2017 for a review). In this paper we focus on the role of IMF orientation (IMF induced).

This paper is a continuation of Tenfjord et al. (2015, 2017), where we studied how a local $B_y$ component is induced in the closed magnetosphere, how the presence of an induced $B_y$ component affects the magnetosphere-ionosphere coupling (Reistad et al., 2016; Tenfjord et al., 2015), and on what timescales this process occurs (Tenfjord et al., 2017). In the previous work we only considered southward directed IMF $B_z$. The purpose of this paper is to show the magnetospheric effects of IMF $B_y$ during northward directed IMF.

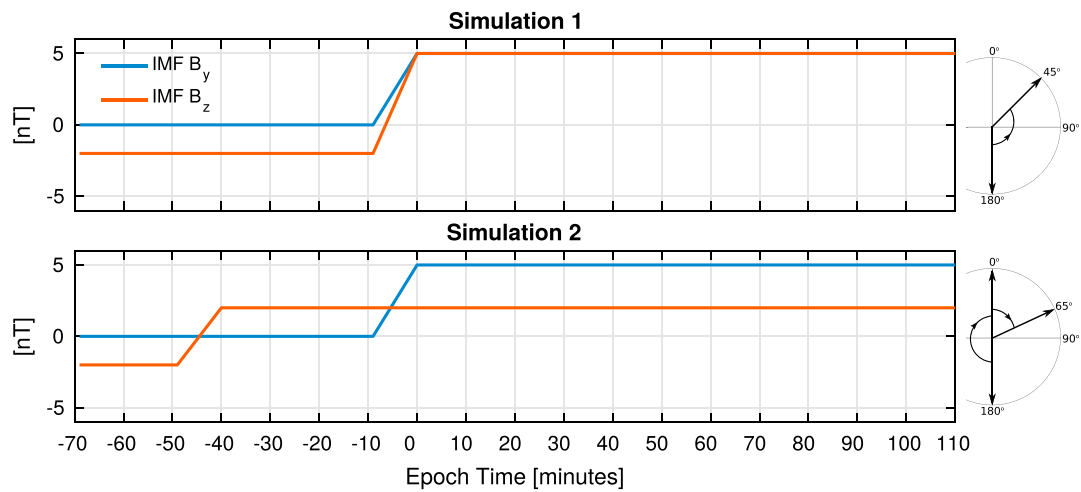

**Figure 1.** Interplanetary magnetic field (IMF) conditions for the two magnetohydrodynamic simulations and illustration of clock angle regimes. Epoch time $= -70$ min corresponds to simulation time $t = 0$.

The outline of the paper is as follows: In section 2 we introduce two simulation runs and the reconnection geometry during northward IMF $B_z$. Section 3 describes the dynamical process which induces the local $B_y$ component. Each step is illustrated using the magnetohydrodynamic (MHD) simulation. In section 4 we study the timescales associated with the induced $B_y$; here empirical and model results are presented. In section 5 our findings are discussed, and our results are compared to earlier studies. Concluding remarks are given in section 6.

## 2. MHD Simulation

We use the high-resolution Lyon-Fedder-Mobarry (LFM) MHD model (Lyon et al., 2004; other models are discussed in section 5.4) to study the evolution of the magnetosphere following an IMF $B_y$ change. We executed two model runs with different IMF conditions (available at ccmc.gsfc.nasa.gov). For both runs the solar wind velocity and density are fixed at $V_x = -450$ km/s and $\rho = 10$ cm$^{-3}$, the temperature is $T = 2 \cdot 10^6$ K, and IMF $B_x = 0$ nT and zero dipole tilt. The different IMF conditions are shown in Figure 1.

In the first simulation the IMF $B_z$ and $B_y$ change at the same time from a purely southward IMF state. In the second simulation, the IMF $B_z$ changes from negative to positive 40 min prior to the $B_y$ component change from 0 to 5 nT. The magnetic field components change from their initial value to their final value in 9 min. Solar magnetic coordinates are used throughout the paper.

Our motivation for running two scenarios is to study the interaction between the IMF and the magnetosphere for different clock angles to investigate how the distribution of the energy loading changes. The reconnection location is controlled roughly by the clock angle, which determines the location of the maximum shear between the IMF and the magnetospheric field. In simulation 2 (65°) the region of maximum shear is shifted further duskward from noon in the Northern Hemisphere compared to simulation 1 (45°; see Komar et al., 2014, and references therein). How this affects the evolution of the energy flow into the magnetosphere is discussed in section 3. Running the two simulations also allows us to compare the magnitude of the induced $B_y$ component for different clock angles.

These two scenarios also enable a study of how the magnetosphere responds to IMF $B_y$ when the magnetosphere has a large polar cap (simulation 1) compared to small polar cap (simulation 2) where the magnetosphere has less stored energy. If the LFM model runs with a fixed northward $B_z$ for a prolonged time, the polar cap becomes very small as a result of closing of the magnetotail field lines (Fedder & Lyon, 1995; Raeder et al., 1995).

### 2.1. Reconnection Geometry

The reconnection geometry for northward IMF has been studied extensively through theoretical consideration (Crooker, 1979; Lockwood & Moen, 1999; Reiff & Burch, 1985; Song & Russell, 1992), spacecraft observations (e.g., Trattner et al., 2004), and MHD simulations (e.g., Komar et al., 2014, Laitinen et al., 2006, Vennerstrom et al., 2005, Watanabe et al., 2004).

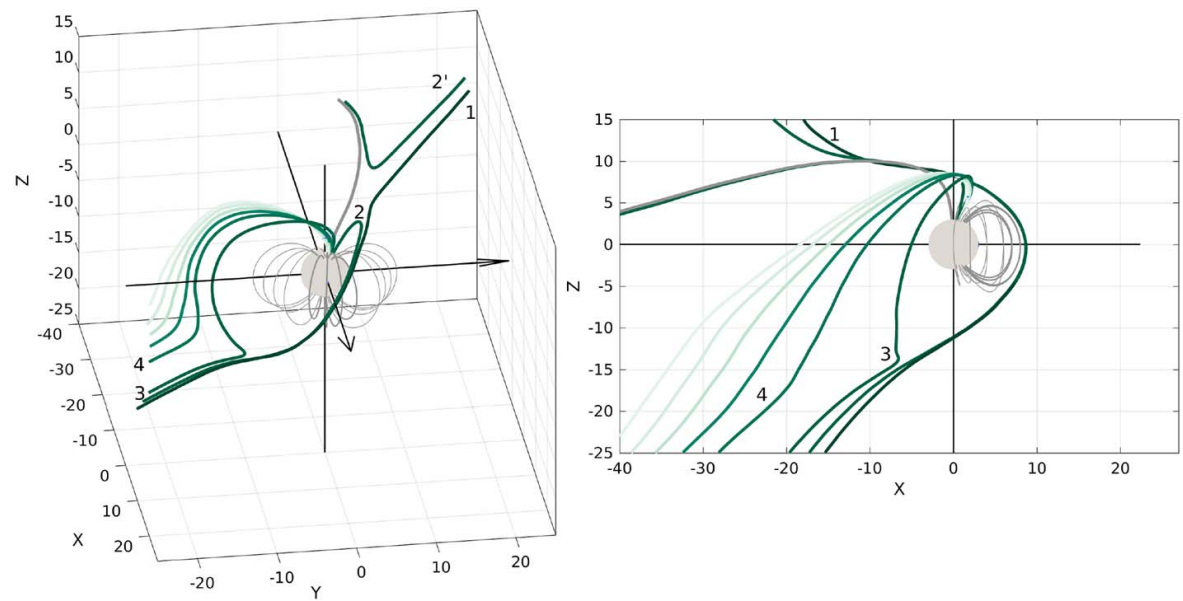

**Figure 2.** Single-lobe reconnection geometry for $\theta = 45°$. Illustration of incoming interplanetary magnetic field (1) reconnecting with the high-latitude lobe field (gray) on the duskside. After reconnecting, one part returns to the solar wind (2′), while the other part (2) is an open field line connected to the Northern Hemisphere. The y-directed magnetic tension on this field line forces the field line toward dawn, while at the same time the solar wind flow forces it tailward. Field lines 3 and 4 represent the evolution of (2).

In Figure 2 the reconnection geometry for a 45° clock angle is shown. The incoming IMF field line reconnects with a magnetospheric lobe field line on the post noonside (originally a part of the duskside lobe). The single-lobe reconnection illustrated produces an open field line connected to the Northern Hemisphere. The field line marked by 2′ returns to the solar wind. The other part (marked by 2) has one foot point in the Northern Hemisphere at high latitudes on the post noonside and is open at the other end. The solar wind carries the field line tailward (marked by 3) and transmits stress along the field line toward the ionosphere. The tension on this newly reconnected field line forces the field line (and the ionospheric foot point) dawnward.

Between points 2, 3, and 4, the field line has been forced dawnward and tailward, and it is wrapped around the magnetopause. When new field lines are continuously being added in its wake (by continuation of dayside reconnection), a pressure arises which forces the flux tubes in the dusk direction to prevent a buildup of flux, thus completing the circulation.

Single-lobe reconnection is also expected in the Southern Hemisphere, where the tension will force the field line toward dusk. This is the reconnection geometry that eventually leads to asymmetric loading/rearranging of flux and asymmetric energy flow. Flux is removed from the northern dusk lobe (2′) and added (2) to the northern dawn. The simulation suggests that both single and dual-lobe reconnection occurs (not shown).

## 3. Generation of Induced $B_y$ in Magnetosphere

The methodology of this analysis is similar to that of Tenfjord et al. (2015) and Tenfjord et al. (2017). We use a global MHD simulation with idealized model input. The results from the model are used to illustrate the mechanism. In this section we describe the dynamical evolution of the process, in which tension on newly reconnected field lines redistributes the open flux asymmetrically between the two hemispheres. This leads to asymmetric magnetic energy density (magnetic pressure), and magnetospheric shear flows that act to equilibrate the dawn/dusk asymmetry will also affect the closed magnetospheric field lines.

To illustrate how the asymmetries in the lobe arise, we have used a slightly different approach compared to that of Tenfjord et al. (2015). Instead of showing the evolution of field lines (advection) and how they are added asymmetrically to the two different lobes, we will here focus on energy flow. The Poynting flux represents the transport of magnetic enthalpy density. For ideal MHD the Poynting vector can be expressed as $\vec{P} = \frac{\vec{v}_\perp B^2}{\mu_0}$, describing that the energy flow is simply the magnetic energy transported by the perpendicular flow (Parker, 2007). Thus, the divergence of the Poynting flux represents energy deposited into the system (Palmroth et al., 2003; Rosenqvist et al., 2008).

To demonstrate the energy flow into the magnetosphere during northward IMF, we show the Poynting flux on a magnetosphere-shaped surface based on the empirical model by Shue et al. (1997). The surface is scaled so that it is smaller than the actual magnetopause. The magnetopause radius changes from about 9 $R_E$ at $X = 0$ and to 13 $R_E$ at $X = -35\ R_E$, while the distance to the subsolar point is approximately 6.5 $R_E$. This is to emphasize the energy flow just inside the system. Thus, it does not represent the total energy transfer from the solar wind to the magnetosphere but rather a fraction of it.

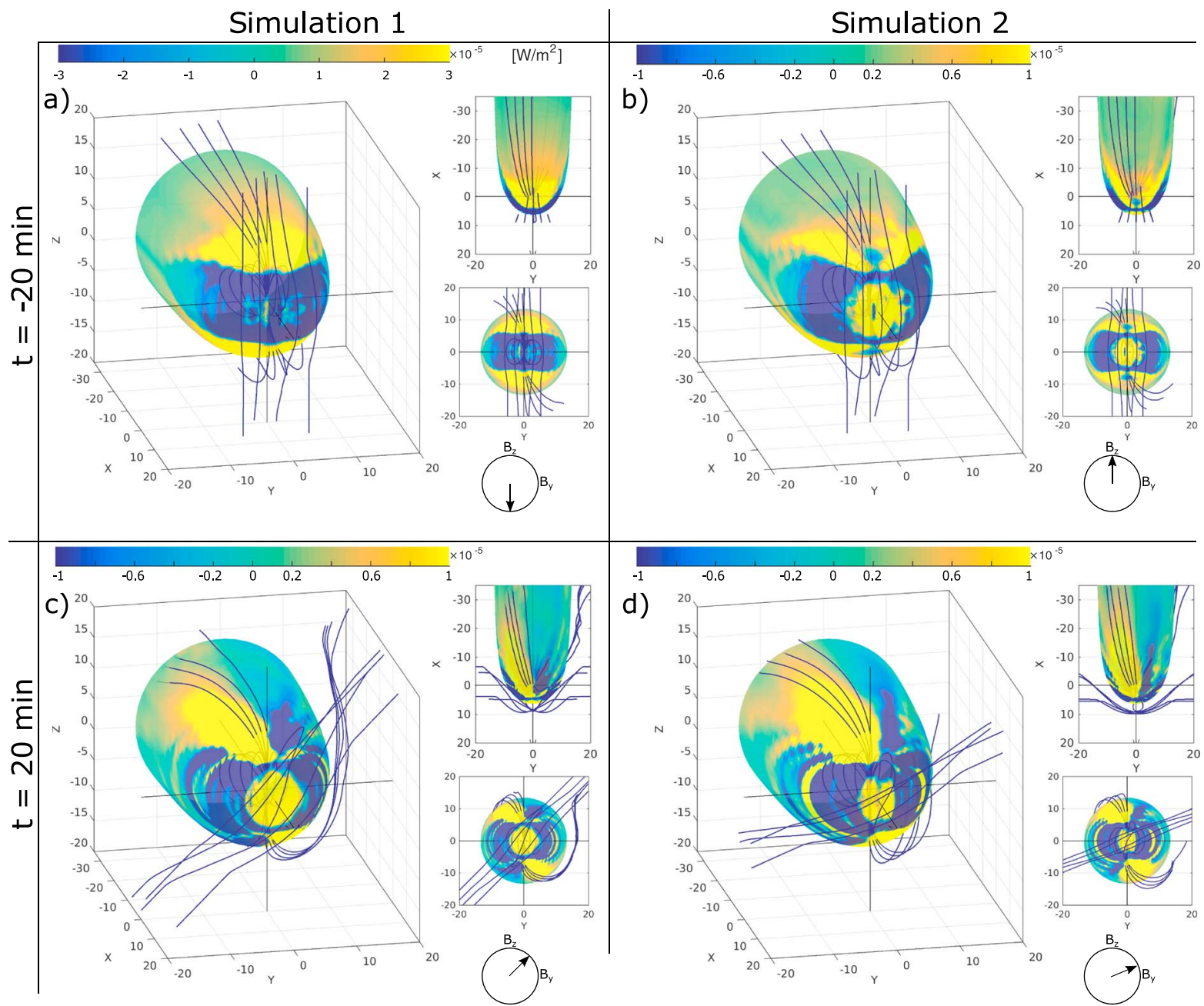

**Figure 3.** Energy flow across an empirical magnetopause-shaped surface located inside the magnetosphere for both simulations. Positive Poynting flux is defined as energy flow into the surface (yellow), and negative (blue) represents energy flow out of the surface. (a, b) The distribution of Poynting flux prior to the arrival of interplanetary magnetic field (IMF) $B_y$. In simulation 1, the IMF $B_z$ is southward and has a much higher energy input compared to simulation 2 (note the different color axis in panel a). (c, d) The distribution of energy flow 20 min after the IMF $B_y$ has arrived.

In Figure 3 we show a selection of IMF, lobe, and closed field lines. The Poynting flux is shown on a magnetopause-shaped surface and represents energy flow into (in yellow) or out of (blue) this surface. For purely northward and southward IMF (Figures 3a and 3b) the energy flow is symmetric about the noon-midnight meridian. In the discussion that follows, we refer to Northern Hemisphere unless stated otherwise. In Figure 3a, reconnection initially occurs between closed magnetic field lines on the dayside and the incoming IMF. The blue regions surrounding the nose of the surface represents the transport of fresh flux from the dayside magnetosphere, out to the magnetopause where reconnection occurs. The reconnected field lines are then dragged across the polar cap and eventually added to the lobes indicated by the yellow regions at high latitude.

In Figure 3b, for purely northward directed IMF (occurring at $-40 < t < -10$), the outward directed (blue) energy flow at high latitude around noon represents the flow of lobe flux refilling the reconnection region with fresh flux. On either side of this region, the energy flow is inward directed, representing reconnected field lines being dragged by the solar wind around the magnetopause nose, while the foot points remain anchored in the Northern Hemisphere. These field lines will circulate toward noon and complete the cycle. The effect on the ionospheric flow are characterized by two symmetric lobe convection cells.

When we introduce IMF $B_y$ (Figures 3c and 3d), the reconnection location shifts from high-latitude noon toward dusk for positive $B_y$ (see the reconnection geometry in Figure 2). The energy flow is now asymmetric. Figures 3c and 3d show that the outward directed energy flow (blue) has shifted duskward from the noon-midnight meridian. Fresh flux is now transmitted from the dusk lobe to the reconnection site. The newly reconnected field lines are forced toward dawn by the tension on the newly reconnected field line. This results in magnetic flux being removed from the dusk lobe and added to the dawn lobe. The effect of this reconnection process is redistribution of magnetic flux from the northern dusk lobe to the northern dawn lobe for positive IMF $B_y$ (opposite for IMF $B_y < 0$). Figures 3c and 3d show the difference in energy flow between 45° (panel c) and 65° (panel d) clock angle. On the low-latitude dayside, the energy flow into the dayside magnetosphere is greater for the 45° clock angle compared to the 65° run. We believe that this reflects a higher level of dual-lobe reconnection for the smaller clock angle. Thus, in Figure 3c, more of the incoming flux reconnects in both hemispheres simultaneously. This results in open lobe field lines being converted to closed field lines. These newly reconnected field lines are convected into the magnetosphere resulting in a positive energy flow at low latitudes as seen in Figure 3. The distribution of energy flow in the tail lobes is also different between the two clock angles. The distribution for the 45° clock angle run (panel c) is shifted slightly more toward lower latitudes compared to the 65° run (panel d). This can be explained by the reconnection geometry or more precisely as the convection of the open field line after reconnection. As seen in Figure 2, the open end of the field lines are carried tailward by solar wind. As the other end is anchored to the Northern Hemisphere, these field lines will eventually be forced duskward, as they straighten out, or experience a magnetic pressure due to the continued loading of magnetic flux to that region. However, the larger the clock angle becomes, the farther northward these will penetrate the magnetopause. We note that there is an additional effect related to the difference in IMF $B_T = \sqrt{B_y^2 + B_z^2}$ between the two simulations. Simulation 1 has a larger magnetic field which results in a smaller magnetopause. The surfaces in Figure 3 are equal in size, which means that the surface in simulation 1 is closer to the actual magnetopause. Also visible in the lower panels are Kelvin-Helmholtz waves at the flanks. These waves are a source of momentum transport across the magnetopause (e.g., Nykyri et al., 2017, and references therein); however, they do not affect our regions of interest.

### 3.1. Asymmetric Magnetic Energy Density

Figure 4 shows the perturbation energy density (or magnetic pressure) on the same surface as in Figure 3. The perturbation magnetic energy density is defined by subtracting the energy density distribution prior to the arrival of IMF $B_y$ ($t = -20$ min) from the energy density at $t = +20$ min: $\Delta U = \frac{1}{2\mu_0}(B_{\text{after}}^2 - B_{\text{prior}}^2)$. We are only showing the result from simulation 2 since the quiet time in simulation 1 has a southward directed IMF $B_z$, making it difficult to extract the perturbation magnetic energy density arising from $B_y$ effects alone. Since the Poynting flux represents the transport of magnetic enthalpy, we expect to see a greater magnetic energy density where the energy flow is positive. Figure 4 shows that the northern dawn and the southern dusk lobes experience an increase in energy density, while in the opposite lobes the energy density has decreased. This illustrates that similarly to IMF $B_z < 0$, the magnetic flux is added asymmetrically (compare to Figure 1 in Tenfjord et al., 2015) to the two lobes in the presence of IMF $B_y$ for IMF $B_z > 0$.

### 3.2. Induced Shear Flows

As seen in Figure 4 the perturbation magnetic energy density extends all the way from the high-latitude lobe down to the central magnetotail (middle panel). The consequence of the enhanced magnetic pressure in the northern dawn (combined with the reduction in northern dusk) results in shear flows, as dictated by the momentum equation. The evolution of $v_y$, considering electromagnetic forces alone, is (cf. Tenfjord et al., 2015)

$$\rho \frac{dv_y}{dt} = \frac{1}{\mu_0}\left(B_x \frac{\partial}{\partial x} + B_z \frac{\partial}{\partial z}\right)B_y - \frac{1}{2\mu_0}\frac{\partial}{\partial y}\left(B_x^2 + B_z^2\right) \quad (1)$$

In Figure 5 we show the $y$ component of the perpendicular velocity $v_{\perp y}$ as color coded and the transverse perpendicular velocity, $v_t = (v_{\perp y}, v_{\perp z})$, as arrows. The plane is located at $X = -15\ R_E$, and we are looking from the tail toward the Earth. For both simulations, we present the perpendicular flows at three different times. At $t = -16$ min the IMF is oriented purely southward in the top panel and northward in the lower panel. At this time there is little convection in the lobes, but for both cases the velocity converges toward the center, which we believe is a signature of reconnection or return flow. At $t = 12$ min, the IMF $B_y$ has initiated asymmetric loading of magnetic flux to the northern dawn and southern dusk. For both simulations we observe shear flows emerging from these regions of enhanced magnetic energy density. At $t = 40$ min the $y$-directed flow is greatly enhanced in the northern dawn and southern dusk. In simulation 2 these flow patterns very much resemble the model presented by Khurana et al. (1996, their Figure 5) and in Liou & Newell (2010, their Figure 3). The asymmetric $y$-directed flows in the lobe induces a local $B_y$ component, described by the induction equation (see Tenfjord et al., 2015, equations (3) and (4)). The white lines at $t = 12$ and $t = 40$ depict the approximate boundary between open and closed field lines, found by field line tracing. The asymmetric loading of flux forces the closed field line region into the negative $z$ direction in dawn and oppositely in dusk. This results in the twisting of the magnetotail (Cowley, 1981; Kaymaz et al., 1994; Petrukovich, 2009, 2011; Walker et al., 1999). The signature of tail reconnection is clearly seen in the figures; close to the magnetic equator the flows exhibit complex patterns as a consequence of reconnection and the associated return flow. Flux is transported across the separatrix feeding the reconnection process, and the resulting outflow can dominate over the asymmetric flows from IMF $B_y$. This feature is prominent in simulation 1 at $t = 12$ min. Regardless of the complex flow patterns from the reconnection process, shear flows that develop as a consequence of asymmetric loading of flux are superposed on to the existing pattern and create the induced $B_y$.

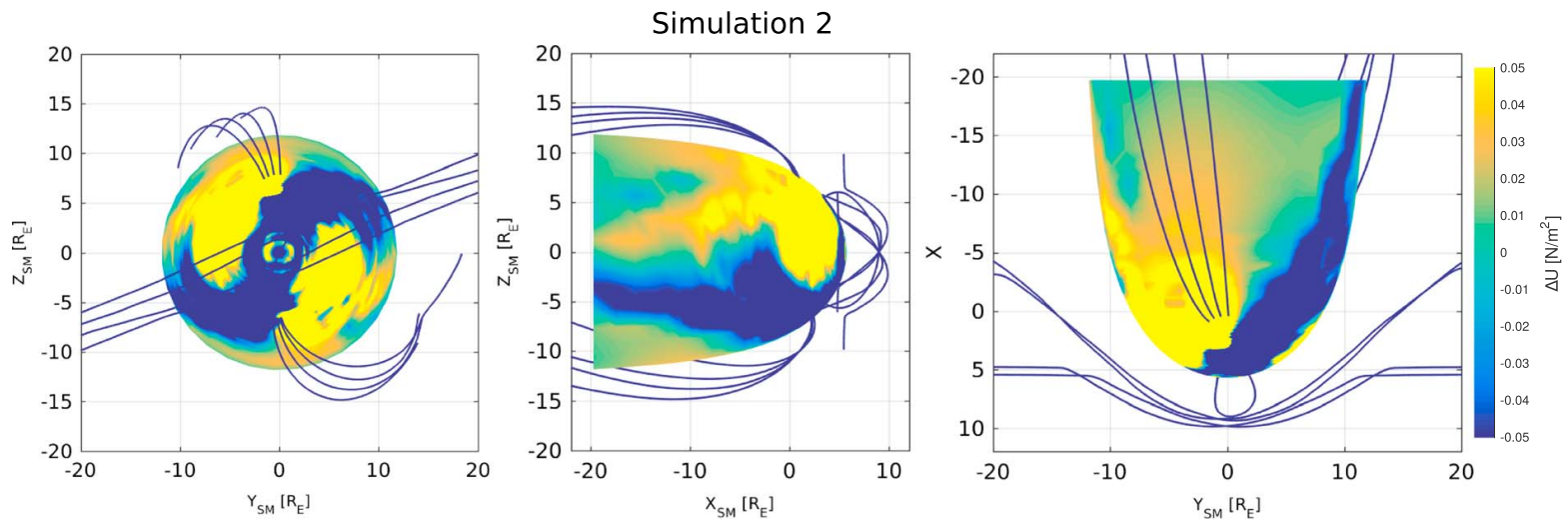

**Figure 4.** Perturbation magnetic energy density from simulation 2, plotted on the same surface as Figure 3. The quiet time subtracted from the current magnetic energy density represents the distribution prior to the arrival of interplanetary magnetic field $B_y$: $\Delta U = \frac{1}{2\mu_0}(B^2_{after} - B^2_{prior})$. The magnetic energy density increases in the northern dawn and southern dusk regions, while it decreases in the northern dusk and southern dawn due to interplanetary magnetic field $B_y$.

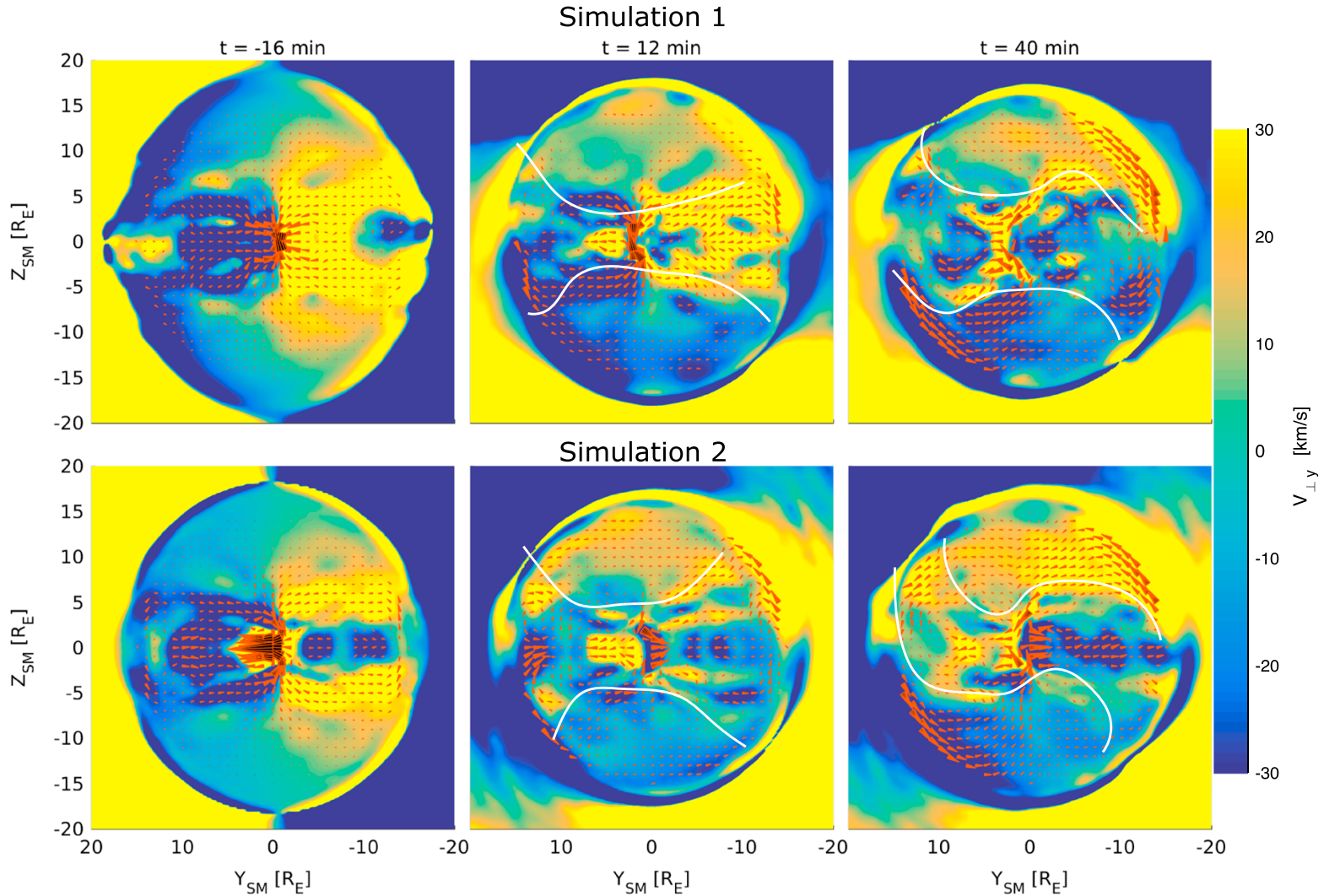

**Figure 5.** Slice plane at $X = -15\ R_E$ showing the $y$ component of the perpendicular velocity $v_{\perp y}$ in color and $v_t = (v_{\perp y}, v_{\perp z})$ as arrows. The shear flow emerges from the northern dawn and southern dusk, corresponding to the regions of enhanced magnetic energy density. White lines illustrate the boundary between open and closed field lines. See text for details.

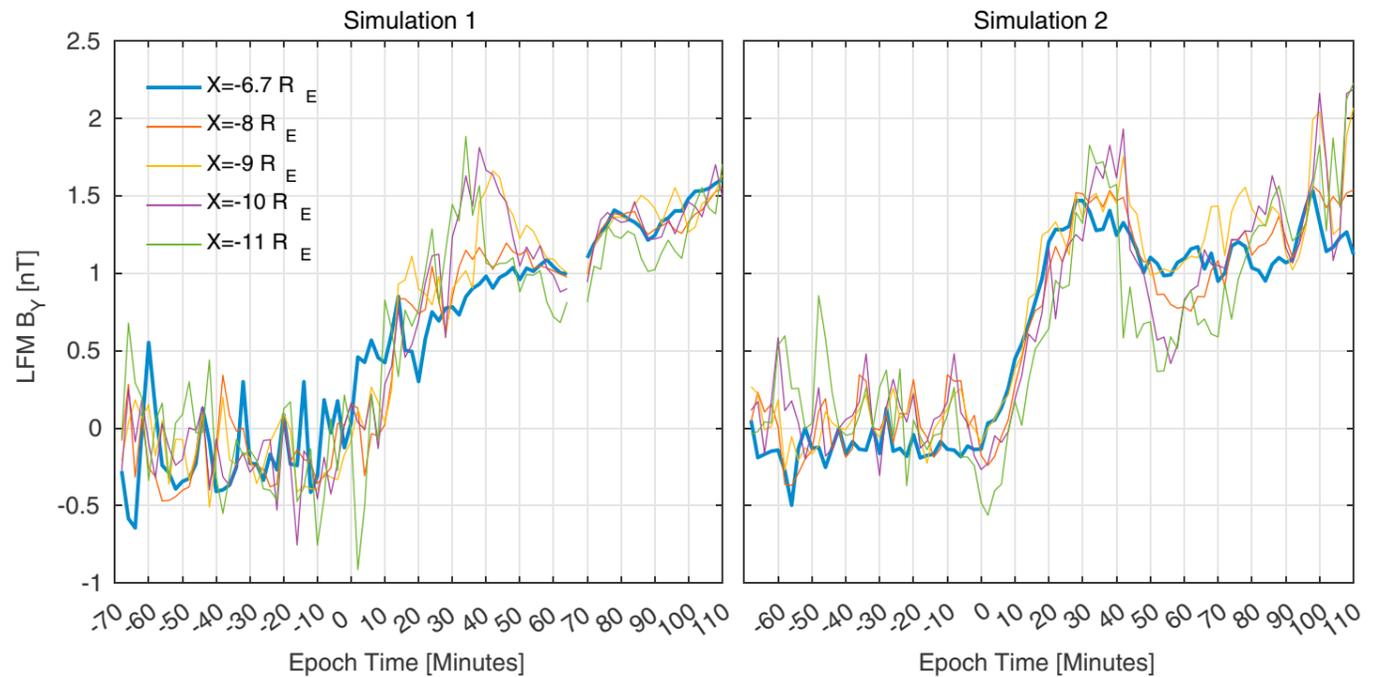

**Figure 6.** Radial dependence of induced $B_y$ on closed field lines for both simulations. The values are taken at $Y = Z = 0$.

To summarize, both simulations clearly illustrate the induced shear flows arising from the asymmetric loading of flux. The patterns are more laminar in simulation 2, which is probably because simulation 1 started out with a negative IMF $B_z$ and therefore contains more stored magnetic energy.

The process by which IMF $B_y$ induces a local $B_y$ component during northward directed IMF is therefore similar to the scenario where the IMF is southward. For both cases asymmetric energy loading into the lobes leads to asymmetric magnetic energy density, which results in oppositely $y$-directed flows between the Northern and Southern Hemispheres. These shear flows effectively produce a $B_y$ component on the already present field lines. The immediate difference between southward and northward IMF conditions is the dayside reconnection geometry.

How these flow patterns and the magnitude of the lobe velocities compare to observations will be discussed in section 5.

### 3.3. Radial Dependence of $B_y$

It has previously been reported that the induction efficiency (discussed in section 5.3) of $B_y$ is higher during IMF $B_z > 0$ compared to $B_z < 0$ (e.g., Cowley & Hughes, 1983). The magnitude of the local $B_y$ component depends on radial distance. As will be discussed below, this is perhaps even more true when $B_z$ is positive than when it is negative. In Figure 6 we show a set of time series of simulation $B_y$ values at five different radial distances. The blue line is the $B_y$ component at geosynchronous orbit, which we will compare to measurements in section 4. We again notice that simulation 1 contains more fluctuations compared to simulation 2, which again, is probably due to a higher level of magnetotail activity due to southward directed IMF $B_z$ prior to the arrival of IMF $B_y$. The response time, defined as the time between the change of IMF $B_y$ at the bowshock and the onset of a change in the local (internal) field, for both simulations, is approximately 10 min, although it is difficult to determine exactly in simulation 1. There does not appear to be a significant difference in the response time between the different locations. The end of the reconfiguration time, as seen in simulation 2, occurs at approximately $t = 30$, which means that it takes approximately 40 min to completely reconfigure (taken into account that the IMF $B_y$ starts to change at $t = -9$ min). These results will be compared to observations in section 4.

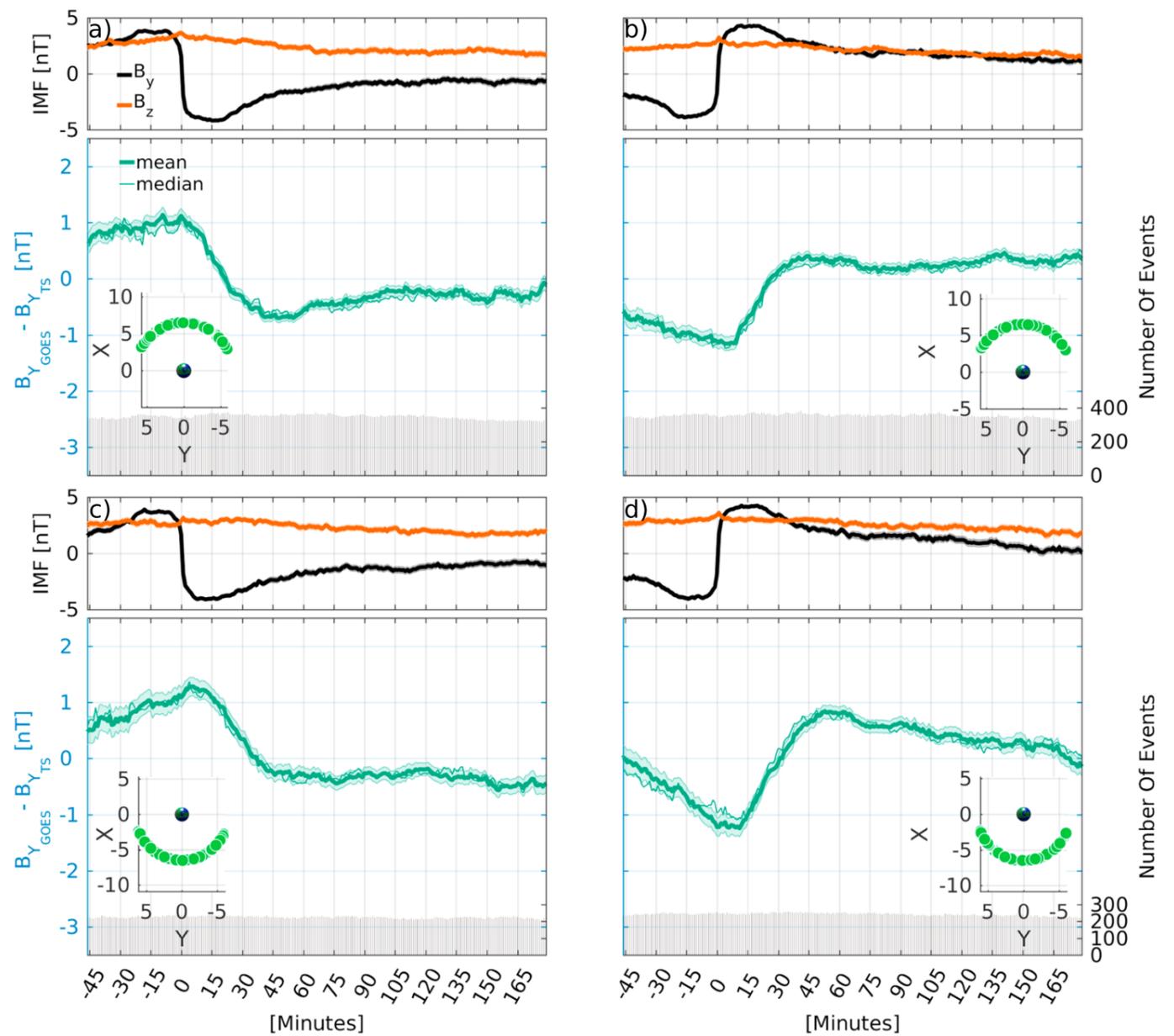

**Figure 7.** (a, b) Averaged interplanetary magnetic field (IMF) $B_y$ and $B_z$ with shaded error region corresponding to standard error of the mean. (c, d) The mean and median $B_y$ component measured by Geostationary Operational Environmental Satellite (GOES) with the TS01 model field subtracted. Bar plots at the bottom show number of events. The x axis represents minutes from epoch. The embedded figures show the location of GOES at $t = 0$.

In simulation 2 between $t = 10$ and $t = 40$ min there is a trend of $B_y$ being induced quicker at smaller radial distances. By comparing $B_y$ at geosynchronous distance (blue) with the value at $X = -11\,R_E$ (green), there appears to be persistent delay of about 5 min. This is consistent with the mechanism described above. The induced $B_y$ is a result of the induced shear flows arising from the asymmetric loading of flux. The flux will be added closer to Earth prior to being added to the lobes farther downtail. If $B_y$ was transported into the closed magnetosphere through tail reconnection, as suggested by, for example, Stenbaek-Nielsen and Otto (1997), we would expect to see the response and reconfiguration first at $X = -11\,R_E$. Pitkänen et al. (2016) reported that the magnetotail farther out responds first to variations in IMF $B_y$. These arguments, and the trend seen in Figure 6, suggests that this is not the case. The same trend is difficult to observe in simulation 1 due to fluctuations. In both simulations, the induced $B_y$ component increases in magnitude with increasing radial distance. This suggests that the shear flows also increase with increasing distance. Comparing the y-directed perpendicular flow in a volume $-6 > y < 6, 10 > z < 4\,R_E$ between $-4 > x > -8\,R_E$ and $-8 > x > -11\,R_E$, we find the velocity to be ~18% higher in the latter region.

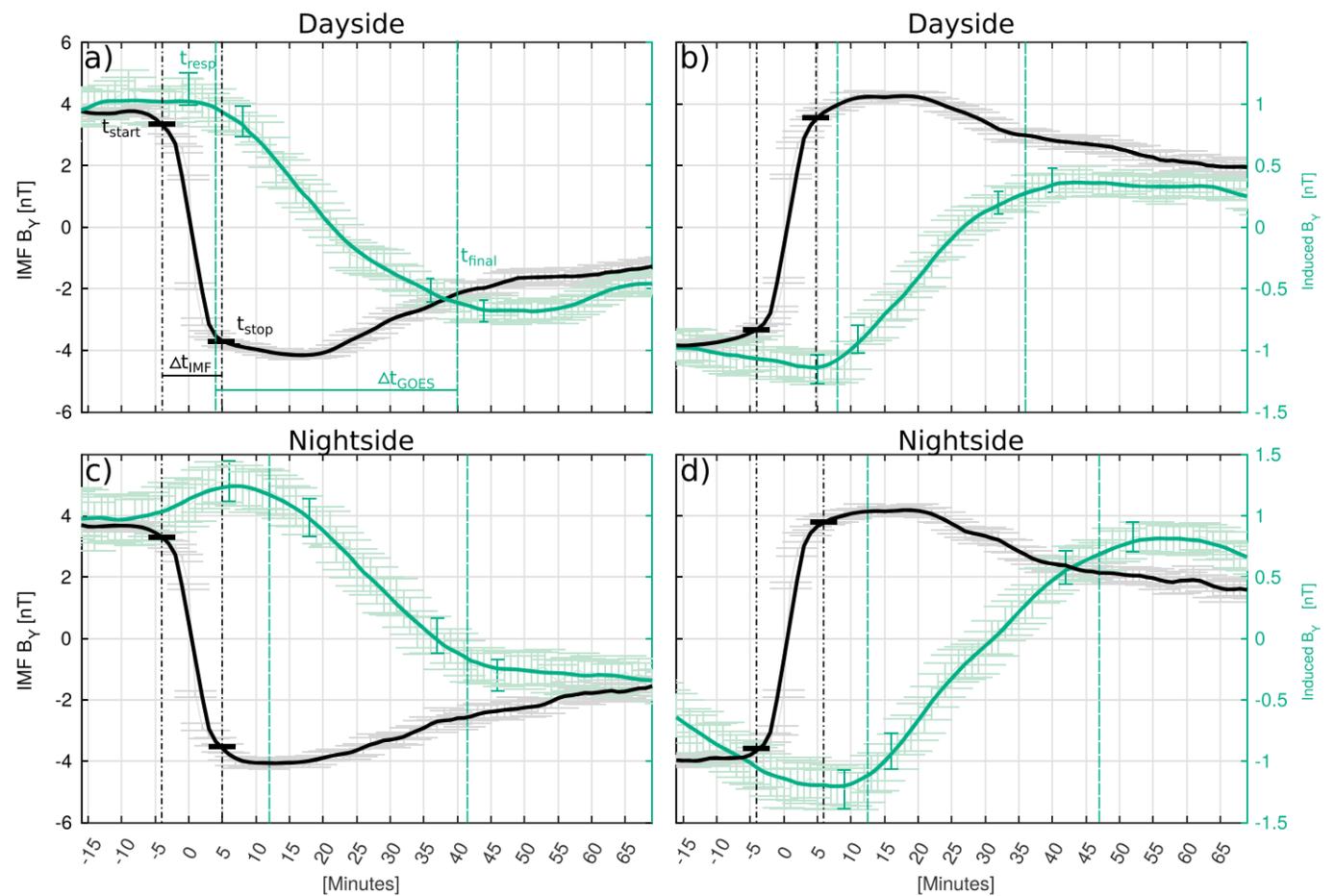

**Figure 8.** Interplanetary magnetic field (IMF) $B_y$ (black) and Geostationary Operational Environmental Satellite (GOES) $B_y$ (green) correspond to Figures 7a and 7b and Figures 7c and 7d but are smoothed using a running mean with a 10-min step length. Black vertical dash-dotted lines show the defined $t_{start}$, the time when IMF $B_y$ begins the reversal, and $t_{stop}$ indicates the time when IMF $B_y$ ends its rotation. The green vertical dashed lines show the defined GOES response and the GOES final time. The uncertainty in GOES $B_y$ represents the standard error of mean. See text for details and Table 1 for values. (a, b) Dayside. (c, d) Nightside.

## 4. Magnetospheric Response to IMF $B_y$ Reversal

In this section we determine the response and reconfiguration times using a large number of IMF $B_y$ reversals and data from Geostationary Operational Environmental Satellite (GOES) at geosynchronous orbit. The analysis is based on magnetospheric data from GOES 8, 9, 10, 11, 12, 14, and 15 and OMNI data from 1997 to 2017. We identify IMF $B_y$ reversals, both positive to negative and vice versa. The IMF $B_y$ is required to be < −2 nT prior to the reversal and > 2 nT after the reversal for negative-to-positive reversals (and reversed for the opposite reversal). Also, events where the $B_y$ changes polarity within 1 hr after the initial reversal are discarded. The methodology is similar to that used in Tenfjord et al. (2017), and we refer to that paper for a detailed description. In Tenfjord et al. (2017) the response and reconfiguration times were found for IMF $B_z < 0$ conditions. Strict criteria had to be set in order to isolate the IMF-induced mechanism from other mechanisms and to avoid magnetic perturbations related to magnetotail activity. For northward directed IMF the magnetotail activity is generally small. In addition we have added GOES 14 and GOES 15 data and updated our list of events to now span from 1997 to 2017. In Figure 7 we show a superposed epoch analysis between the IMF $B_y$ reversals and magnetic field measurements from seven GOES spacecraft in the dayside and nightside. The Tsyganenko model (TS01; Tsyganenko, 2002) magnetic field with IMF $B_y = 0$ is subtracted from the measured $B_y$. Figures 7a and 7b show the IMF conditions for each reversal at each location. The IMF $\Delta B_y$ is approximately 8 nT for all panels, about equally distributed prior and after the reversal. Figures 7c and 7d show the measured GOES $B_y$ component where the Tsyganenko model field with IMF $B_y = 0$ is subtracted. Extreme events, where the IMF, solar wind velocity, or density exceed the values where the Tsyganenko model is valid, have been removed from the database. For both reversals and locations, Figure 7 shows a prompt response, followed by a gradual reconfiguration spanning over approximately 30 min. This is similar to the results for IMF $B_z < 0$ shown in Tenfjord et al. (2017).

**Table 1**
*IMF $B_y$ and GOES $B_y$ Response and Reconfiguration Times Given in Minutes*

| | IMF | | | | GOES | | | Ratio |
|---|---|---|---|---|---|---|---|---|
| | Epoch start | Start | Stop | | Response | Final | | $\Delta t_{GOES}/$ |
| | $t_e$ (min) | $t_{start}$ | $t_{stop}$ | $\Delta t_{IMF}$ | $t_{resp}$ | $t_{final}$ | $\Delta t_{GOES}$ | $\Delta t_{IMF}$ |
| Dayside +/− $B_y$ | −4 ± 2 | 0 ± 2 | 9 ± 2 | 9 ± 2.8 | 8 ± 4 | 44 ± 4 | 36 ± 5.5 | 4.0 ± 0.35 |
| Dayside −/+ $B_y$ | −4 ± 2 | 0 ± 2 | 9 ± 2 | 9 ± 2.8 | 12 ± 3 | 40 ± 4 | 28 ± 5 | 3.1 ± 0.36 |
| Nightside +/− $B_y$ | −4 ± 2 | 0 ± 2 | 9 ± 2 | 9 ± 2.8 | 16 ± 6 | 41.5 ± 4.5 | 25.5 ± 7.5 | 2.8 ± 0.43 |
| Nightside −/+ $B_y$ | −4 ± 2 | 0 ± 2 | 10 ± 2 | 10 ± 2.8 | 12.5 ± 3.5 | 47 ± 6 | 34.5 ± 7 | 3.45 ± 0.35 |

*Note.* All times are relative to the bow shock reference time. Note that there is an additional 4–8 min for the IMF phase fronts to propagate from the bow shock to the magnetopause. IMF = interplanetary magnetic field; GOES = Geostationary Operational Environmental Satellite.

### 4.1. Characteristic Response and Reconfiguration Times

To estimate the response and final reconfiguration times, we use the same technique as in Tenfjord et al. (2017). To determine the time between IMF $B_y$ reversals ($t_{start}$) and the corresponding $B_y$ response at GOES ($t_{resp}$), we filter the data using a running mean with a 10-min step length (see Figure 8). By visual inspection we identify the first signature of changes in $B_y$. This becomes our lower bound. The upper bound is determined by the standard error of mean, by identifying the first value outside the uncertainty of our lower bound as illustrated in Figure 8a. The same method has been applied to determine $t_{final}$ and IMF $t_{stop}$. Table 1 summarizes the observed response and reconfiguration time and their uncertainties. Note that approximately 5 min should be subtracted from the GOES response and reconfiguration time to account for the propagation from the bow shock to the subsolar point. The response and reconfiguration times from the presumed arrival at the magnetopause are presented in parentheses in the following section.

From Table 1 we conclude that the magnetospheric response time, at all local time positions, is less than 16 (11) min from the bow shock (magnetopause) arrival time. In less than 47 (42) min the magnetospheric state has reached its final configuration ($t_{final}$). $\Delta t_{IMF}$ and $\Delta t_{GOES}$ are defined as the time between the beginning and end of the reversals. The ratio $\Delta t_{GOES}/\Delta t_{IMF}$ describes the relationship between the slope of IMF $B_y$ and the magnetospheric $B_y$; thus, it describes how quickly the magnetosphere reconfigures with respect to the time it takes IMF $B_y$ to rotate. From Table 1 we can see that it takes 3–4 times longer for the magnetosphere to reconfigure to match the external forcing.

This shows that $B_y$ is induced during northward IMF $B_z$ on timescales that are very similar to those found during southward directed IMF. They are also comparable to the timescales found in the simulation LFM $t_{resp} \sim 10$ min and LFM $t_{final} \sim 40$ min. This strengthens our conclusion that the mechanism inducing $B_y$ is similar; asymmetric loading of flux launches compressional waves giving rise to the fast response, and the gradual reconfiguration time follows from the magnetosphere adjusting into the new asymmetric state. The magnitude of the induced $B_y$ is smaller at all local times (compare to Figures 4 and 5 in Tenfjord et al., 2017); this will be discussed in section 5.3.

## 5. Discussion

The objective of this study has been to explain each step of the mechanism which eventually leads to an induced $B_y$ component in the magnetosphere. In addition, we have characterized the response and reconfiguration times from both data and simulations. In this section we will discuss the consequences of our findings and how these compare to earlier work.

### 5.1. IMF Magnitude Dependence

The magnitude of the induced $B_y$ is usually related linearly to the magnitude of IMF $B_y$. This proportionality is usually called "penetration efficiency": induced $B_y = \epsilon \cdot IMFB_y$, where $\epsilon$ has been found statistically to vary between 0.1 and 0.8 (Fairfield, 1979; Kaymaz et al., 1994; Petrukovich, 2009; Tenfjord et al., 2017).

We here propose an alternative scaling hypothesis based on simple arguments, arguing that this may not simply have a linear relationship but instead scales with a coupling function between the solar wind and magnetosphere.

The energy transfer from the solar wind into the magnetosphere depends in general on the upstream velocity, density, the transverse magnetic field $B_T (= \sqrt{B_y^2 + B_z^2})$, and the clock angle ($a\tan(B_z, B_y)$). In empirical and theoretical coupling functions the dependency on $B_T$ varies generally with a power between 2/3 and 2 (Du et al., 2011; Gonzalez, 1990; Newell et al., 2008; Tenfjord & Østgaard, 2013; Vasyliunas et al., 1982). A general coupling function can be thought of as a combination of energy flux and efficiency. The efficiency is usually incorporated in the clock angle dependency, accounting for the length of the reconnection line and the efficiency of the dayside reconnection. The amount of energy flux able to enter the magnetosphere depends on this coupling efficiency. The energy transferred into the magnetosphere is extracted from the solar wind mechanical energy. The mechanical energy flux is converted to electromagnetic energy as newly reconnected field lines are dragged tailward by the solar wind. Thus, it is in fact not the magnetic energy density in the solar wind that is the direct source of magnetospheric energy density. Its role is instead to control dayside reconnection and transport of energy via the deformation of the newly reconnected field lines. However, the mechanical and electromagnetic energy fluxes are coupled, since the force extracting linear momentum from the solar wind flow is electromagnetic ($\vec{J} \times \vec{B}$). The energy extracted from the magnetosheath flow is thus the work done by the Lorentz force on the magnetosheath flow. Let us for simplicity assume that the coupling function is given simply by $P_C = v_{sw} B_T^2 F(\theta)$, where $v_{sw}$ is the solar wind speed, $B_T$ is the transverse IMF field, and $F(\theta)$ represents the efficiency through a clock angle and/or a effective length dependency.

We suggest that the induced $B_y$ component depends on the induced shear flow $v_y$, which originates from a gradient of the perturbed lobe magnetic field $\delta B$. Furthermore, the relation between the magnetic perturbation inside the magnetosphere scales as an energy coupling function to the upstream solar wind conditions. This connects the perturbation magnetic field $\delta B$ to $v_{sw} B_T^2 F(\theta)$, where $\delta B$ is the perturbation to the magnetic lobe energy density when a small amount of newly reconnected magnetic flux is pushed through the magnetopause by the solar wind. The resulting $\delta B$ gives rise to a velocity. The integrated $y$ component of the momentum equation, assuming cold plasma and steady state, is given as $\rho v_y^2 = \frac{1}{2\mu_0}\left[(B_0 + \delta B)^2 - B_0^2\right]$. The first term on the right-hand side represents the total magnetic field given as a sum of the unperturbed field ($B_0$, which is primarily $B_x$ here) and the perturbed field ($\delta B$). Thus, the induced shear flow produced by the perturbation field can then be expressed as $v_y = \sqrt{\frac{1}{2\mu_0 \rho} B_0 \delta B} = V_A \sqrt{\frac{\delta B}{2}}$. It is the shear flow that induces the local $B_y$ component through the induction equation, which in the assumption of incompressible flow can be expressed as $\frac{dB_y}{dt} = (B_x \frac{\partial}{\partial x} + B_y \frac{\partial}{\partial y} + B_z \frac{\partial}{\partial z}) v_y$. Assuming that the gradient of $v_y$ along the $z$ direction dominates, $B_y$ can be expressed as $B_y = B_z \frac{v_y}{L_z} t$. Thus, the induced $B_y \propto v_y$, while $v_y \propto B_0 \delta B \propto v_{sw} B_T^2 F(\theta)$.

To test our hypothesis, we used our GOES database to study two IMF $B_y$ reversals with different IMF $\Delta B_y$ but with similar average clock angle (~51°), solar wind velocity, and density. The first case had IMF $\Delta B_y = 5.4$ nT (symmetric such that IMF $B_y = \pm 2.7$ nT) and IMF $B_z = 2.3$ nT, and the second had IMF $\Delta B_y = 11.9$ and IMF $B_z = 4.6$ nT. The induced $\Delta B_y$ was 1.3 nT for the first case and 3.9 nT for the second. Which means that while the strength of the IMF $B_y$ increased by a factor of 2, the induced $B_y$ increased by a factor of 3. Our hypothesis is that the induced $B_y$ should be proportional to the IMF $B_T^2$. The ratio between the induced $B_y$ is 3, while the ratio between $B_T^2$ in the two cases is 4.5. This suggests that induced $B_y$ scales with IMF $B_T^{3/2}$.

Tenfjord et al. (2017) used GOES to investigate the coupling efficiency during southward directed IMF $B_z$. The authors grouped the data into velocity and IMF $B_y$ subsets. They found that larger velocity and stronger IMF $B_y$ resulted in higher efficiency, but they did not consider the IMF $B_T$ component.

For both our simulations the strength of the IMF $B_y$ was fixed at 5 nT. The IMF $B_z$ component is 5 nT in simulation 1, compared to 2 nT in simulation 2. As seen in Figure 6 the magnitude of the induced $B_y$ is similar in the two simulations, even though $B_T$ is ≈30% higher in simulation 1. This enhancement in $B_T$ in simulation 1 may have been countered by the smaller clock angle resulting in a less efficient distribution of energy flow, as seen in Figure 3.

The main point of this discussion is to show that there is no theoretical argument for IMF $B_y$ to be linearly correlated to the induced $B_y$. The reason for the existence of this comparison arises from the idea that the IMF $B_y$ "penetrates" the magnetosphere, and the local $B_y$ component is determined by the superposition principle. Instead, we suggest that the magnitude of the induced $B_y$ component depends on the energy input to the system, which determines the strength of the magnetic energy density, and the location to which the additional flux is added is controlled by the clock angle.

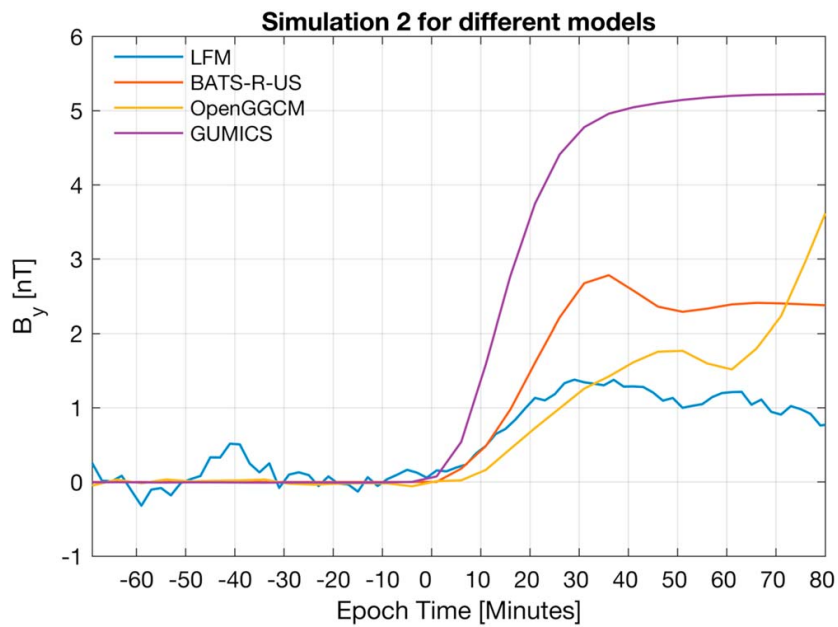

**Figure 9.** Comparing the response and reconfiguration time of the induced $B_y$ at $X = -6.7 R_E$ between Lyon-Fedder-Mobarry, Block-Adaptive-Tree-Solarwind-Roe-Upwind-Scheme (BATS-R-US), Open Geospace General Circulation Model, and Grand Unified Magnetosphere-Ionosphere Coupling Simulation (GUMICS). The values are the taken at $Y = Z = 0$.

### 5.2. Magnitude of Shear Flows

The shear flows at $X = -15 R_E$, shown in Figure 5, can be compared to Haaland et al. (2008, 2009), where Cluster Electron Drift Instrument was used to map the convection velocity to a $YZ$ plane at $X = -10 R_E$. Haaland et al. (2008) conservatively defined the lobe as an $8 \times 16 R_E$ region centered at $Y = 0$, $Z = 12 R_E$ and similarly in the Southern Hemisphere. Comparing Figure 4 in Haaland et al. (2008) with our Figure 5, there is a very good correspondence between the overall flow patterns. The magnitude of the $v_{\perp y}$ in the lobes is larger than what Haaland et al. (2008) reported. Their results are from $X = -10 R_E$, while the flow patterns shown in Figure 5 are from $X = -15 R_E$. We find that the velocities are larger at $X = -15$, but in the following comparison we extract the flow velocities at $X = -10 R_E$ from the simulations. In the northern lobe, for southward directed IMF, Haaland et al. (2008) found a average $v_z$ velocity of $-10$ km/s. In the simulation this velocity is $v_z = -14$ km/s. For northward directed IMF the authors reported $V_z = -2$ km/s, compared to LFM $V_z = -1$ km/s. For positive IMF $B_y$ with clock angle sector around $90° \pm 22.5°$ the authors found $V_y = 6.5$ km/s compared to LFM $V_y = 11$ km/s taken at $t = 12$ min (lower right panel in Figure 5) or $V_y = 18$ km/s at $t = 40$ min (middle panel, simulation 2). While the $z$-directed velocities are comparable between the simulation and the statistical Cluster study, the $V_y$ velocity is considerably larger in the simulation. One explanation could be that this discrepancy is due to different clock angles, $\theta = 90° \pm 22.5°$ in Haaland et al. (2008), compared to $\theta = 65°$ in our simulation).

### 5.3. Induction Efficiency

Although we argue that the induced $B_y$ may not depend linearly on IMF $B_y$, a comparison between the efficiency from previous studies is in order. For both the nightside and dayside, and for both types of reversals, the IMF $\Delta B_y = 8$ nT (Figure 7 or 8). For the dayside the induction efficiency is about 21% and 25% for panels (a) and (b), respectively. For the nightside we found 19% and 25% for panels (c) and (d). However, in section 5.1 we analyzed the induced $B_y$ for two different cases with equal clock angle. For the strongest solar wind driver, the coupling efficiency was 33% on the nightside (negative-to-positive reversal). These induction efficiencies are weaker compared to southward IMF, for which we found on average a 30% coupling efficiency on the dayside and ∼50% on the nightside (Tenfjord et al., 2017).

To our knowledge, the only statistical analysis of local $B_y$ response to IMF $B_y$ during northward IMF has been done by Cowley and Hughes (1983). The authors presented the penetration efficiency at four quadrants: dawn, dusk, dayside, and midnight. The maximum efficiency was found around midnight and the lowest in the dusk quadrant (all measurements at geosynchronous orbit). Our results are obtained as an average over the entire nightside or dayside (Figure 7). Around midnight Cowley and Hughes (1983) found an efficiency of 54%, and around noon of 19%. Their average over all quadrants gave an efficiency of 33%. Their coupling efficiency at midnight is higher than what we observe. One possible explanation is that we have used the TS01 model as our background field, compared to Cowley and Hughes (1983) where nine standard daily field variations were calculated to act as baseline values. Even though we use IMF $B_y = 0$ in the model input, the TS01 model has incorporated other processes for generating $B_y$ which may influence the efficiency when compared to these results. Cowley and Hughes (1983) also observed a trend in the data toward increasing induced $B_y$ for decreasing negative IMF $B_z$. The authors found that for strong southward IMF $B_z$ the induction efficiency became similar to that of northward IMF $B_z$. This nonlinear behavior may be related to our discussion in section 5.1.

### 5.4. MHD Model Comparison

In this section we compare the LFM global MHD model to the Open Geospace General Circulation Model (OpenGGCM; Raeder et al., 1998), BATS-R-US (Gombosi et al., 2003; Powell et al., 1999), and GUMICS-4 (Janhunen et al., 2012).

We only show the $B_y$ evolution at $X = -6.7 R_E$ in the midnight ecliptic plane $(Y = 0, Z = 0)$. In Figure 9 we show the response and reconfiguration times and the magnitude of the induced $B_y$ in the three listed models above at $X = -6.7 R_E$ as a function of time.

The three additional simulations show a prompt response comparable to LFM (Figure 6) and to our empirically obtained response (Figure 7). GUMICS and BATS-R-US shows a comparable reconfiguration time to LFM, while OpenGGCM predicts a slightly longer reconfiguration time. The magnitude of the induced $B_y$ in GUMICS is 3–4 times larger compared to LFM. The magnitude of the induced $B_y$ predicted by LFM appears to be closest to the observed values at this distance (Figure 7). Although there are differences between the four models, such as integration scheme, spatial and temporal resolution, and inclusion of coupled ionosphere/plasmasphere models, they all tell the same quantitative story, thus giving confidence in the result. A more detailed discussion of exact evolution and amplitudes is beyond the scope of this paper.

## 6. Summary

In this paper we presented a dynamical description of how the IMF $B_y$ induces a local $B_y$ component in the magnetosphere during northward IMF conditions. We have described each step and used the LFM MHD model to illustrate the dynamical process. The response and reconfiguration times has been determined by both modeling and by empirical GOES data. There is a good compliance between the empirical and simulated timescales. Our key findings are listed below:

1. Merging between northward directed IMF with $B_y$ and terrestrial field results in asymmetric loading/redistribution of magnetic flux due to tension on newly reconnected field lines.
2. In the magnetospheric lobes this results in asymmetric magnetic energy density between the dawn and dusk lobes.
3. Shear flows are induced as to restore magnetic equilibrium (primarily $V_y$).
4. Through the induction equation, these shear flows induce a local $B_y$ component.
5. The magnetosphere responds to the IMF $B_y$ component after 8–16 min at geosynchronous distances, with respect to the bow shock arrival time.
6. The reconfiguration time is 40–47 min for all local times.
7. Both the response and reconfiguration times are remarkably similar to the southward IMF case shown in Tenfjord et al. (2017).
8. The ratio between induced $B_y$ and IMF $B_y$ is smaller during northward IMF compared to southward IMF at geosynchronous distances.
9. LFM MHD simulation suggests that the induced $B_y$ increases with radial distances on the nightside, which we suggest is due to the distribution of shear flows in the magnetotail.
10. We predict that the coupling efficiency between IMF $B_y$ and local $B_y$ not only depends simply linearly on IMF $B_y$ but scales as an energy coupling function.
11. From theoretical arguments and MHD simulations we predict the local $B_y$ to respond faster in the magnetotail at smaller radial distances.

**Acknowledgments**
This study was supported by the Research Council of Norway/CoE under contract 223252/F50. S. E. Milan was supported by STFC grant ST/K001000/1. We acknowledge the use of NASA/GSFCs Space Physics Data Facility for OMNI data. Simulation results have been provided by the Community Coordinated Modeling Center at Goddard Space Flight Center through their public Runs on Request system (http://ccmc.gsfc.nasa.gov). The CCMC is a multiagency partnership between NASA, AFMC, AFOSR, AFRL, AFWA, NOAA, NSF, and ONR. LFM runs are available as Paul_Tenfjord_081117_1 and Paul_Tenfjord_073117_1a, BATS-R-US: ilja_honkonen_011918_1, OpenGGCM: ilja_honkonen_011918_3, and GUMICS: ilja_honkonen_011918_4.